\documentstyle[12pt]{article}
\begin{document}
\rightline{IMSc-94/25}
\vglue .2in
\centerline{\tenbf A LOOK AT THE DISCRETIZED SUPERSTRING}
\baselineskip=22pt
\centerline{\tenbf USING RANDOM MATRICES\footnote{To appear in the
Proceedings of the II International Colloquium on Modern Quantum Field
Theory, Bombay, India, January 5-11, 1994.} }
\vspace{0.8cm}
\centerline{\tenrm PARTHASARATHI MAJUMDAR}
\baselineskip=13pt
\centerline{\tenit The Institute of Mathematical Sciences}
\baselineskip=12pt
\centerline{\tenit Madras 600113, India}
\newcommand{\be}{\begin{equation}} \newcommand{\ee}{\end{equation}}
\newcommand{\bea}{\begin{eqnarray}} \newcommand{\eea}{\end{eqnarray}}
\vspace{1.9cm}
\abstract{Beginning with a review of the arguments leading to the
so-called c=1 barrier in the continuum formulation of noncritical
string theory, the pathology is then exhibited in a discretized
version of the theory, formulated through dynamical triangulation of
two dimensional random surfaces. The effect of embedding the string in
a superspace with fermionic coordinates is next studied in some
detail. Using techniques borrowed from the theory of random matrices,
indirect arguments are presented to establish that such an embedding
may stabilize the two dimensional world sheet against
degeneration into a branched polymer-like structure, thereby leading
to a well-defined continuum string theory in a spacetime of dimension
larger than 2.
}

\vfil
\baselineskip=14pt
\section{The $c_m=1$ Barrier}

\subsection{Continuum Formulation}

Bosonic string theory in twenty six dimensional spacetime is
characterized by two important properties : (i) quantum fluctuations
of the string world sheet metric (Liouville mode) decouple, leaving
behind a free theory of matter and ghost fields in two dimensions;
(ii) the theory has exact two dimensional conformal symmetry. The
first property is no longer true for embeddings of the string in
spacetimes of dimensionality $D \neq 26$, thus requiring a proper
quantum formulation of the Liouville mode. It is however possible to
deal with this within the 2d conformal field theory framework provided
the 2d cosmological constant is treated perturbatively. In this
formulation, the Liouville mode behaves like a free scalar field with
a background charge $Q$, such that its contribution to the central
charge of the Virasoro algebra is given by $c_L=1+3Q^2$. Requiring now
that the total central charge, consisting of contributions from the
matter, ghost and the Liouville sectors, vanish fixes $Q$ to be
\begin{equation}
Q~=~\left ( {{25~-~c_m} \over 3} \right )^{\frac12}  \label{bkgd}
\end{equation}
The fixed area partition function scales as
\be
Z(\lambda A)~~=~~\lambda^{({Q\over \lambda} -1)} Z(A)~~, \label{fixas}
\ee
where, $\gamma \equiv -\frac12 Q + \left ( {1-c_M \over 12} \right
)^{\frac12} $.
Clearly, $Q/{\gamma}$ is complex for $1 \leq c_M < 25$, so that the
string susceptibility has complex critical exponents for $c_M$ in
this interval. This is the so-called $c_M=1$ barrier\cite{{kpz},{ddk}}.
If we interpret the Liouville mode as an extra dimension of
spacetime \cite{das1}, then the interval $2 < D <26$ is the forbidden
region for the bosonic string. What exactly happens to string dynamics
for these dimensions of the embedding space is not clearly understood;
the malady has been variously attributed to tachyons in the string
spectrum\cite{sei}, a strongly coupled phase of two dimensional gravity or to
the disintegration of the string world sheet to branched
polymers.

A tachyon-free string theory is of course one with target space
supersymmetry which allows naturally for spacetime fermions. The
continuum Green Schwarz superstring\cite{gsw}, which is classically
consistent in spacetime dimensions 3, 4, 6 and 10 \cite{sie} is an
important candidate because, in addition to the usual bosonic string
coordinates $X^{\mu}(\sigma, \tau)~,~\mu=1, \dots D-1$ it also has spinorial
coordinates $\theta^{\alpha}(\sigma, \tau)~,~\alpha=1, \dots
2^{(D-1)/2}$, which are spinors in target space but (anticommuting)
scalars on the world sheet much like the $b-c$ ghost system of the
bosonic string. Now recall that the latter always contribute
negatively to the total conformal anomaly (central charge). If the
fermionic coordinates indeed have a similar dynamics, then quite
conceivably $c_{\theta}<0$, so that, with $c_X=D-1$ and
$c_M=c_X+c_{\theta}$, one has $D > c_M+1$. This implies that the
$c_M=1$ barrier no longer restricts the allowed dimensionality of
spacetime. While it is known that for critical $D=10$ Green Schwarz
superstrings this is indeed true, the situation for $D \neq 10$ is far
from clear. The major problem has to do with the fermionic gauge
symmetry known as $\kappa$ symmetry\cite{sie} and its Lorentz
covariant fixing\cite{maj1}. This, however, is not relevant for
the discretized version of the model formulated as a dynamically
triangulated world sheet embedded in superspace\cite{amb1}. We next
turn to this version of the theory.

\subsection{Discretized Formulation}

The discretized bosonic string \cite{{amb2},{dav},{kaz}} is
given, for a random world sheet of spherical topology, by
\be
Z_B(\beta, \Lambda)~~=~~\sum_T {{e^{- \Lambda |T|}} \over \rho(T)}
Z_B^T(\beta)
\ee
where,
\be
Z_B^T (\beta)~\equiv~\prod^{D-1}_{\mu=1} \int \prod_{i=1}^V d X^{\mu}
\prod_{<ij>} dP^{\mu}_{ij} e^{-\beta S_B(X,P)}~~, \label{bosp}
\ee
and the action, expressed in first order form, is given as a sum over
links by
\be
S_B(X,P)~\equiv~\sum_{<ij>} \left[ \frac12 (P^{\mu}_{ij})^2 + i
P^{\mu}_{ij} (X^{\mu}_i - X^{\mu}_j) \right] ~~. \label{bosac} \ee
The variables $P_{ij}$ are basically link variables on the triangular
lattice, which we define to be antisymmetric under interchange of the
indices $i$ and $j$. Integrating over these variables yields the more
familiar version of the discretized Polyakov string whose action is
\be
S_{Polyakov}~~\sim~~\sum_{<ij>} (X_i - X_j)^2~~. \ee

If we scale
\be
X_i^{\mu} \rightarrow \beta^{-1/2} X_{i}^{\mu}~;~ P^{\mu}_{ij}
\rightarrow \beta^{-\frac12} P^{\mu}_{ij}~~, \ee
then
\be
Z_B^T(\beta)~=~\beta^{-\frac12 (D-1)(|T| + V - 3)} Z_B^T(1)~~\label{betas}
\ee
which implies that $Z_B(\beta, \Lambda)$ can be thought of as a power
series in $\beta^{-1}$. For a fixed genus of the world sheet, the
total number of triangles into which the world sheet has actually been
triangulated, $|T|$, increases, with the number of vertices $V
\rightarrow \infty$, as  $[c(g)]^V$, where $c(g)$ is a number of order
1. Therefore there is a large $\beta = \beta_0^B$ such that
\be Z_B(\beta) < \infty ~~for \beta > \beta^B_0~~, \label{zbeta} \ee
and diverges as $(\beta - \beta_0)^{\alpha}$ at criticality.
Correspondingly, the string susceptibility $\chi_B (\beta)$ diverges
at criticality with exponent $\Gamma$.

The pathology inherent in the existence of a forbidden range of
embedding space dimension manifests in the discretized version in the
critical behaviour of the lattice string tension. For the existence of
a proper continuum limit, the lattice string tension should scale to
zero at criticality. But both for hypercubic and triangular
latticizations of the random world sheet, there exists a non-zero absolute
lower bound on the lattice string tension, proportional to the
critical temperature (inverse string coupling) \cite{ambd} for $D>2$.
Although
it is very likely that random surfaces of minimal area (spikes)
dominate the partition function, leading to a degeneration of the
world sheet into branched polymers, the only evidence comes from numerical
simulations which show this behaviour for $D>11$ \cite{amb3}. Another
aspect of this malady manifests in the Hausdorff dimension
$$d_H~\equiv~ \lim_{|T| \rightarrow \infty} <X^2>_{|T|}/ln
{|T|}~~.$$
For $D > 2$, numerical studies indicate that
$$<X^2>_{|T|}~~\sim~~|T|^{\frac12}~~$$
so that $d_H \rightarrow \infty$. While reflection positivity is a
good working hypothesis ensuring absence of tachyons from
the spectrum, the above sickness might well originate from tachyons in
the continuum. Thus, spacetime supersymmetry is a likely cure.
Additional motivation comes from the theory of random walks: for
supersymmetric walks the Hausdorff dimension is 1 compared to 2 for
bosonic walks.

\section{Discretized superstring in $D=3$}

\subsection{Scaling properties}

The partition function for a discretized superstring in a superspace
with 3 bosonic Euclidean directions is given by\cite{{mik},{amb1}},
\be
Z_S^T(\beta)~=~\int \prod_{\mu=1}^3 \prod_{i=1}^{|V(T)|} d
X^{\mu}_i \prod_{<ij>} dP^{\mu}_{ij} \prod_{i, \alpha}d
\theta^{\alpha}_i e^{-\beta S_S} ~~, \label{suprtf} \ee
where, the action, in a first order form, is
\be
S_S~\equiv~\sum_{\mu, <ij>} \{ \frac12 (P^{\mu}_{ij})^2 ~+~i
P^{\mu}_{ij} [X^{\mu}_i - X^{\mu}_j + \frac12 i \theta_{[i}
\sigma^{\mu} \theta_{j]}] \}~~. \label{sact} \ee
Upon integration over the momentum variables $P_{ij}^{\mu}$, one
obtains the conventional discretized version of the Green Schwarz
action {\it sans} the Wess-Zumino type term needed for the action to
be $\kappa$ invariant. We recall that the latter is not realized on
the lattice\cite{amb1}; further, in $D=3$ the WZ term is not relevant
even in the continuum beacuse one does not need it for $\kappa$
invariance. One further remark is that, unlike the second order form
in which the fermionic coordinates appear through quartic couplings,
in the first order form in (\ref{sact}) they appear quadratically just
like the $X$s; thus they can be integrated out to yield an effective
action as a functional of the link variables alone.

Performing a scaling of the variables in (\ref{suprtf})
\be
X^{\mu}_i \rightarrow \beta^{-\frac12} X^{\mu}_i~,~P^{\mu}_{ij}
\rightarrow \beta^{-\frac12} P^{\mu}_{ij}~,~\theta^{\alpha}_i
\rightarrow \beta^{-\frac14} \theta_i^{\alpha}~~\nonumber \ee
we get
\be
Z_S^T(\beta)~=~\beta^{-\frac32 [{|T|} +V-3]} \cdot \beta^{-V}
Z_S^T(1) ~~.  \label{scas} \ee
As a power series in $\beta^{-1}$, $Z_S(\beta)$ has a radius of
convergence given by $(\beta^S_0)^{-1}$. A crucial question is, if we
assume that the absolute lower bound on the lattice string tension
derived by Ambjorn and Durhuus in ref. \cite{ambd} in terms of the
critical temperature is still valid, then is $\beta_0^S < \beta_0^B$ ?
If so, the minimal value of the critical lattice string tension will
be smaller for the superstring than for the bosonic string signifying
that supersymmetry is a step in the right direction. Evidence for the
latter is already available in the work of Ambjorn and Varsted who
present numerical results on the average ratio of the radius to
circumference of the random surface\cite{amb1}. These authors show
that this ratio is closer to 1 for the superstring than for the
bosonic string, although the results are inconclusive. In the sequel
we present an analytical approach to this problem. We show that
$\beta_0^S~~<\beta_0^B~~$, under some assumptions. Work is in progress
to ascertain whether the critical string tension is indeed zero.

\subsection{Representation as a Matrix Model : bosonic string}

Even though the first order action (\ref{sact}) is quadratic in the
$\theta$ and $X$ variables, integrating over them yields a rather
complicated action which, in the literature has only been dealt with
numerically. In this subsection we use the theory of random matrices
towards an analytical approach to this problem. Before turning to the
case of the superstring, we take a closer look at the bosonic string
to illustrate our approach. Recall that for a bosonic string in
$D-1=3$ spatial dimensions, the partition function is given, after
integration over $X$ as\cite{amb1}
\be
Z_B^T(1)~=~\prod_{\mu=1}^3 \int \prod_{<ij>} dP^{\mu}_{ij}
e^{-\sum_{<ij>} (P^{\mu}_{ij})^2} \prod_i \delta(\sum_j
P_{ij}^{\mu}) ~~,\label{bospf} \ee
On the other hand, if we do the $P$ integration followed by
integration over the $X$ variables, we get
\be
Z_B^T(1)~~\sim~~det^{-3/2} {\bf D}_{ij}~~, \label{adjcy}\ee
where ${\bf D}_{ij}$ is the adjacency matrix,
$$ {\bf D}_{ij}~\equiv~\left\{ \begin{array}{lll}
                -1 & \mbox{ if $i$ and $j$ are nearest neighbours} \\
                 0 & \mbox{otherwise}
             \end{array}
\right \} $$
With this definition,
\bea
Z_B^T~&=&~\int \prod_{\mu;<{ij}>} dP^{\mu}_{ij} e^{
\sum_{\mu;<{ij}>}  (P^{\mu}_{ij})^2 } \nonumber \\
& \cdot & \{ \prod_{\mu;i} \delta (\sum_j
P^{\mu}_{ij}) \prod_{\mu;ij} \delta ~(( 1-{\bf D}_{ij})P^{\mu}_{ij}) \}~,
\label{bfp} \eea
where, the terms above in curly brackets are constraints that enforce
that the integration variables are in reality link variables.

We now make that assumption that, on the above constraint surface,
link fluctuations in different directions are independent, and as
such, the antisymmetric matrices $P^{\mu}$ commute for different
$\mu$. This assumption now allows us to cast eqn (\ref{bfp}) in the
form of a matrix model with `non-singlet' delta function constraints
for three antisymmetric matrices. If ${\cal O}_{ij}$ is the orthogonal
transformation that reduces each $P^{\mu}$ to its Jordan canonical
form with eigenvalues $p^{\mu}_i$, and $\Delta(p)$ is the Van der Monde
determinant, we have
\bea
Z_B^{T, matr}(1)~&=&~\prod_{\mu} \int \prod_i dp^{\mu}_i e^{\sum_i(p^{\mu})^2}
\nonumber \\
{}~ & &~\Delta^2(p^{\mu}) \int {\cal D} {\cal O} \{ constraints \}[{\cal O}]~
\label{bpmm} \eea
If we ignore the non-singlet constraints,
\be
Z_B^T(1) ~~=~~ [ V! det {\cal H}]^3~~, \ee
where ${\cal H}$ is the Hadamard matrix\cite{bess}
$$ {\cal H }_{ij}~\equiv~\left\{ \begin{array}{lll}
                {(i+j-1)!! \over 2^{\frac12 (i+j)}} & \mbox{ if $i+j$
is even} \\
                 0 & \mbox{$i+j$ odd, $i,j=0,1, \dots V-1$. } \end{array}
\right \} $$
The rationale for considering the case when the non-singlet
constraints are absent is simply that these constraints are the same
for the bosonic and the supersymmetric cases, and as such, the ratio
of the partition functions in their absence is expected to be a good
approximation to the ratio when these constraints are included.

\subsection{Matrix Model Representation for the Discretized
Superstring}

Integrating over the $X$ and $\theta$ variables we get
\begin{equation}
Z_S^T(1)~=~\int \prod_{\mu;<{ij}>} dP^{\mu}_{ij}
e^{-\sum_{\mu;<{ij}>} (P^{\mu}_{ij})^2 } Det
\left(\sum_{\mu}(\sigma^{\mu} P^{\mu})
\right ) \prod_{\mu;i} \delta(\sum_j P^{\mu}_{ij} )~~, \label{sef}
\end{equation}
where $\sigma^{\mu}$ are the Pauli matrices. Using elementary matrix
theory this can be rewritten
\begin{equation}
{\tilde Z}_m^T~=~\int \prod_{\mu;<{ij}>} dP^{\mu}_{ij}
e^{\{-\sum_{\mu;<{ij}>} (P^{\mu}_{ij})^2
- ln(\sum_{\mu}(P^{\mu})^2) \}} \cdot {constraints}~~, \label{sep}
\end{equation}
where, {\it constraints} refer to the delta function constraints in
the rhs of eq. (\ref{bpmm}). Once again, we have a (constrained)
matrix model of three antisymmetric matrices with a potential
which is a sum of Gaussian and logarithmic parts, the latter being due
to the fermionic coordinates. In terms of eigenvalues $p^{\mu}_i$,
this is reexpressed as
\bea
Z_S^T(1)~=~&\int & \prod dp^{\mu}_i e^{-\sum (p^{\mu}_i)^2 }
{}~ \prod_i\left( \sum_{\mu} (p^{\mu}_i)^2 \right) \nonumber \\
&\cdot & \prod \Delta^2(p^{\mu}) \int constraints ~~\label{sepm} \eea

Let us now define a real symmetric matrix of order $V \times
V$ in the following manner :
\be A_d~~\equiv~~diag(a_1, a_2, \dots a_V)~~,~ a_i real~, \ee
\be A~~\equiv {\cal O} A_d {\cal O}^{-1}~~; \label{dixt} \ee
Thus, the matrix $A$ shares its angular parts with those of the random matrices
$P^{\mu}$, while its eigenvalues $\{ a_i \}$ are classical, since they are not
integrated over. $A$ is therefore a rather special type of external
matrix field. Consider now the partition function
\be
{\cal Z}(a_1, \dots a_V)~\equiv~\prod_{\mu} \int \prod_i dp^{\mu}_i
\Delta^2(p) \cdot {\cal C}~~; \label{dad} \ee
here ${\cal C} \equiv \int constraints$, with the constraints being
once again the delta function constraints. Observe that ${\cal C}$ is
quite independent of the eigenvalues $\{ a_i \}$, so that one obtains
the following results \cite{maj2},
\bea
Z^T_B(1) ~~&=&~{\cal Z}(a_1,\dots,a_V) |_{a_1=a_2=\dots=a_V=1} \\
Z^T_S(1)~~&=&~~\prod_i {\partial{} \over \partial{a_i}}
{\cal Z}|_{a_1=a_2=\dots=a_V=1} ~~. \label{rel} \eea
In fact, if we were to expand ${\cal Z}(\{a_i\})$ around the point
$a_i=1~ for~ all~ i$, the bosonic and supersymmetric partition functions
become coefficients of the first and $V$th terms in this expansion.
Thus, the theory described by the partition function ${\cal Z}$ is
interesting on its own right, especially if any of
the other coefficients in the above expansion could be identified as
partition functions of some new string theories.

Once again, if the non-singlet constraints are ignored,
\bea
{\cal Z}^{matr}(a_i)~&=&~\prod_{\mu} \int {\cal D} P^{\mu} e^{-Tr
[A\sum(P^{\mu})^2 ]} \nonumber \\
{}~~&\equiv&~~\prod_{\mu} {\cal Z}_{\mu}~~\label{dadm}, \eea
where ${\cal Z}_{\mu}$ describes a random (antisymmetric) matrix model
with a `diagonal' external field $A$. Recall that for $A$ being the identity,
the bosonic partition function was exactly calculable a l\'a
Bessis\cite{bess} in terms of the determinant of the Hadamard matrix.
For arbitrary real $a_i$ not necesarily all equal to 1, one can define
a generalized Hadamard matrix ${\tilde {\cal H}}$ whose elements are
defined as
\be {\tilde{\cal H}}_{ij}~~=~~a_{i+1}^{-(i+j+1)/2} {\cal
H}_{ij}~. \label{genh} \ee
In terms of this matrix, the partition function in (\ref{dadm}) can be
expressed as
\be {\cal Z}_{\mu}~~=~~det {\tilde {\cal H}}~~ for~ every~ {\mu} ~~.
\label{mine} \ee
The proof of the above consists of a straightforward generalization of
the proof of Bessis for the case when all the external eigenvalues
$a_i$ are unity, and is not included here\cite{maj3}. It lends itself
to generalization quite freely to the cases of non-Gaussian measures.

Specializing to the case of the discretized superstring imbedded in a
superspace of 3 bosonic dimensions, we have
\be
{\cal Z}^{matr}(a_1, \dots a_V)~=~det^3{\tilde {\cal H}} ~~. \ee
Now, from eqn.s (\ref{rel}), (\ref{dadm}), (\ref{genh}) and (\ref{mine}), one
can derive an upper bound on the ratio of the supersymmetric to the bosonic
partition function {in the absence of the delta function constraints} :
\be
{Z_S^{matr} \over Z_B^{matr}} ~<~ \left(\frac{1}{2^V} { {det \left[{ (i+j+1)!!}
  \over
2^{\frac12 (i+j)} \right] } \over {det \left[ {(i+j-1)!!} \over 2^{\frac12(i+j)
}\right ] }} \right)
^3 ~.\label{ubnd} \ee
For the limit $V \rightarrow \infty$, the ratio on the rhs is certainly less
than 1. If we now assume that
\be
{{Z_S^T(1)} \over {Z_S^{matr}}}~~\sim~~{{Z_B^T(1)} \over {Z_B^{matr}}}~~, \ee
because, as already mentioned, the non-singlet constraints in both cases are
identical, we can infer that
\be
Z_S^T(1)~~< ~~Z_B^T(1)~~, \ee
which implies that $\beta_0^S < \beta_0^B$ . Thus if we believe that analogous
to the bosonic case the critical lattice string tension for the superstring is
bounded from above by a quantity proportional to the inverse critical
temperature, then it follows that this minimal value is lower in the
supersymmetric case as compared to the bosonic case.

Two crucial questions remain at this point : (1) is this minimal value of the
string tension actually zero, as one would like it to be ? (2) Is the
geometrical proof of Ambjorn and Durhuus \cite{ambd} for the bosonic string
generalizable to the case of the superstring, as we have assumed above. We do
not have any answer to the first question at this point. What one requires is
a better technique to deal with matrix models with delta function constraints
which are functions of the `angular' variables left over after diagonalization.
As for the second issue, we argue below that there are reasons to suspect that
the proof in the bosonic case may not generalize.

\subsection{Is $T_{lat}^S \geq 2\beta$ ? }

We first briefly review the proof given in ref.\cite{ambd} for the absolute
lower bound of the lattice string tension in case of the bosonic string.
Consider a closed loop $\gamma_{L,n}$ of length $L$ on the random world sheet,
which has $4n$ vertices with coordinates $X_i~,~i=1,\dots n$. The loop
correlation function for the bosonic string is defined as
\be
G_{\beta}(\gamma_{L,n})~\equiv~\sum_{T \epsilon {\cal T}(n)} \rho(T) \int
\prod _{i \epsilon T/ \partial T} d X_i e^-{\beta \sum_{<ij>} (X_i-X_j)^2}~.
\ee
The lattice string tension is defined in terms of this loop Green's function
as
\be
T ~\equiv ~-\lim_{L \rightarrow \infty} {1 \over L^2} ln G_{\beta}(\gamma_
{L,L})~~.\label{tens} \ee
For fixed L, total number of vertices $|T|$ on the entire world sheet, and
for a fixed loop $\gamma$, one can decompose the discretized superstring action
into a term which may be thought of as the minimum (or saddle point) or
classical part of the action, subject to a fixed boundary which maps into the
fixed loop $\gamma$, and a term which is a function only of the fluctuating
variables, and as such, is independent of $\gamma$ :
\be
S[|T|, \gamma]~=~S_{min}(|T|, \gamma) ~+~S'(T)~~, \label{splt}
\ee
with
\be
S_{min}~~=~~\sum_{<ij>} (X_{0i} - X_{0j} )^2~~. \ee
Here $X_0$ indicate `saddle-point' or classical solutions of the discretized
eqations of motion. Note that, this classical part corresponds to a fixed
triangulation of the world sheet. Consequently, it is a sum of squared
Euclidean distances between points which are vertices of triangles. It
follows from the relation between the area and squared length of each side of
an equilateral triangle that
$$S_{min} \geq 2L^2 ~~~.$$
This implies that
\be
G_{\beta}(\gamma_{L,L})~\leq~ e^{-2 \beta L^2} G'_{\beta}~~, \ee
where $G'_{\beta}$ is independent of $\gamma_{L,L}$. In fact, Ambjorn and
Durhuus have shown that $G'_{\beta} \leq e^{cL}~,~c\geq 0$; thus
\be
G_{\beta}~~\leq~~e^{2 \beta L^2}~~, \label{limt} \ee
so that
\be
T_{lat}^{B}(\beta)~~\geq~~2 \beta~~.\label{ltens} \ee
As $\beta \rightarrow \beta_0$, therefore, the lattice string tension has an
absolute lower bound given by twice the inverse critical temperature.

The question now is whether a similar geometrical result exists for embeddings
of the random surface in a superspace of three Euclidean bosonic dimensions.
Observe that the geometric result sketched above depends crucially on the
interpretability of the classical (minimal) action in terms of a sum of
squared Euclidean lengths. For the superstring, the action in
second order form is given by the square of a `current' which is manifestly
supersymmetric under spacetime supersymmetry transformations:
\be
S_S~~=~~\sum_{<ij>} \left(X_i - X_j  + i\theta_{[i} \sigma \theta_{j]}
\right )^2 ~~. \ee
Now the infinitesimal squared super-invarinat interval on flat
superspace is  given as
$$ d {\cal S}^2~~=~ \{ dx ~+~i \theta \sigma d\theta \}^2 ~~,$$
It is not at all clear whether the finite form of this `superlength' has
anything to do with the superstring action. This is more so
because, unlike ordinary Minkowski space, flat superspace has torsion.
One might of course
inquire as to whether, the effective action obtained upon
integrating over the fermionic coordinates in a first order form, can indeed
be given a geometric interpretation akin to the bosonic situation. This issue
is currently investigation \cite{maj3}. Basically one needs to reformulate the
proof of Ambjorn and Durhuus for the first order form of the bosonic string
action and then attempt a supersymmetric generalization.

\section{Summary and Conclusions}

We have presented evidence (albeit indirect) that the discretized superstring
in a superspace with $D=3$ has better chances of metamorphosing into a well
defined continuum string theory than its bosonic counterpart. A more accurate
calculation of the superstring partition function is necessary, especially one
in which the non-singlet constraints are handled adequately, to show explicitly
that the lattice string tension does indeed scale to zero at criticality.

The generalization of Bessis' approach to the case when a special type of
external matrix field is present whose eigenvlaues are classical and angular
parts are random, can be extended to higher polynomial measures as well. Thus,
if we have a matrix model whose potential is
$$ V(M,A)~=~AM^2~+~gA^2M^4~~, $$
with $M$ being an $N \times N$ hermitian matrix which is random, and $A$ is
a hermitian matrix defined as in eqn. (\ref{dixt}), i.e., its angular part
shares the randomness of $M$ while its eigenvalues remain classical, then
the partition function can be shown to be given by
$$ Z(A)~=~N! det {\tilde {\cal H}}~~,$$
where,
$$ {\tilde {\cal H}}_{ij} \equiv a^{-\frac12 (1+j+1)}_{i+1} {\cal H}_{ij} ~,$$
and \cite{maj3}
$$ {\cal H}_{ij}~\equiv~\int dm e^{-(m^2+gm^4)} m^{(i+j)}~ for~ i+j~even~.$$

Finally, apart from the simplicity of $D=3$ for the choice of
the embedding space for the discretized superstring, there is the expectation
that, if a continuum limit exists then the Liouville mode will play the role of
an extra dimension of spacetime, and may lead to a realistic situation, of
course, one {\it without the full $D=4$ Poincar\'e supersymmetry}. Since it is
quite likely that this continuum theory (if it exists) will not have tachyons
in its spectrum, `space supersymmetry' may turn out to be enough to make
amplitudes finite. If so, the problem of the cosmological constant gets
decoupled from the hitherto unsolved problem of a nonperturbative mechanism
for spacetime supersymmetry breaking. This could have far-reaching
implications for future research in string theory.

\vglue 0.4cm


\section{References}
\begin{thebibliography}{99}
\bibitem {kpz} {V. Knizhnik, A. Polyakov and A. Zamolodchikov, Mod.
Phys. Lett. {\bf A3}, 819 (1988).}
\bibitem {ddk} {F. David, Mod. Phys. Lett. {\bf A3}, 509 (1988); J.
Distler and  H. Kawai, Nucl. Phys. {\bf B321}, 509 (1988). }
\bibitem {das1} {S. Das, A. Dhar, A. Sengupta and S. Wadia, Mod. Phys.
Lett. {\bf A5}, 799 (1990).}
\bibitem {sei} {N. Seiberg, in {\sl Random Surfaces and Quantum Gravity}, ed.
by O. Alvarez et. al., Plenum Press (1991).}
\bibitem {amb1} {J. Ambjorn and S. Varsted, Phys. Lett. {\bf 257B},
305 (1991).}
\bibitem {gsw} {M. Green, J. Schwarz and E. Witten, {\sl Superstring
Theory}, vol. 1,  Cambridge University Press (1988).}
\bibitem {sie} {W. Siegel, {\it Introduction to String Field Theory},
World Scientific (1989).}
\bibitem {maj1} {For a review see P. Majumdar, Aspects of Green Schwarz
Superstrings, in {\it Modern Quantum Field Theory} ed.s S. Das et. al., World
Scientific (1990); see also P. Majumdar, Mod. Phys. Lett. {\bf A5}, 1505
(1990).}
\bibitem {amb2} {J. Ambjorn, B. Durhuus and J. Fr\"olich, Nucl. Phys.
{\bf B257}, 433 (1985). }
\bibitem {dav} {F. David, Nucl. Phys. {\bf B257}, 543 (1985). }
\bibitem {kaz} {V. Kazakov, I. Kostov and A. Migdal, Phys. Lett. {\bf
157B}, 295 (1985).}
\bibitem {amb3} {J. Ambjorn, B. Durhuus, T. Jonsson and G.
Thorleifsson, Niels Bohr Institute Preprint NBI-HE-92-35 (1992). }
\bibitem {ambd} {J. Ambjorn and B. Durhuus, Phys. Lett. {\bf 188B},
253 (1987).}
\bibitem {mik} {A. Mikovi\'c and W. Siegel, Phys. Lett {\bf 240B}, 363
(1990).}
\bibitem {maj2} {P. Majumdar, Phys. Rev. Lett. {\bf 71}, 1140 (1993). }
\bibitem {bess} {D. Bessis, Comm. Math. Phys {\bf 69}, 147 (1979). }
\bibitem {maj3} {P. Majumdar, in preparation.}
\end{thebibliography}
\end{document}